# Pulse duration retrieval using a commercial laser diode with a monitor photodiode

ADRIAN F. CHLEBOWSKI*, LUKASZ A. STERCZEWSKI, JAROSLAW SOTOR

*Laser & Fiber Electronics Group, Faculty of Electronics, Photonics and Microsystems, Wroclaw University of Science and Technology, Wybrzeze Wyspianskiego 27, 50-370 Wroclaw, Poland*
*[adrian.chlebowski@pwr.edu.pl,](mailto:adrian.chlebowski@pwr.edu.pl)*

**Abstract:** Semiconductor lasers merge coherent light emission with photodetection and, owing to third-order nonlinearities in their active region, function as sensitive room-temperature two-photon absorption (TPA) detectors. Here, we leverage these capabilities offered by a commercially available InGaAsP semiconductor laser diode with an integrated InGaAs monitor photodiode to measure the duration of femtosecond pulses at telecom wavelengths. The nonlinear and linear signals obtained from both detectors enable the retrieval of pulse duration with femtosecond accuracy without moving parts. This capability is demonstrated down to 2 µW average optical power for a 250 MHz repetition rate laser.

## 1. Introduction

Femtosecond laser pulses are indispensable in a wide range of applications, including spectroscopy [1], precision metrology [2], material processing [3], or biomedical imaging [4]. Direct optoelectronic detection at these timescales is not feasible; therefore, advanced diagnostic methods are required for characterizing repetitive ultrafast light pulses. Since the inception of the celebrated intensity autocorrelation (IAC) technique [5], various implementations have employed a moving delay arm to probe time-averaged pulse characteristics. Not only could pulse duration be assessed, but also, in more sophisticated implementations, the temporal pulse profile [6,7].

Conventional techniques, such as IAC, require one to produce a delayed pulse replica, which spatially overlapped with the original pulse illuminate a nonlinear crystal or a suitable nonlinear photodetector to generate a nonlinear correlation signal. Since this is often achieved via mechanically scanning the delay, the robustness and compactness of this well-established approach are rather limited. Alternatives that eliminate mechanical scanning map delay onto a spatial coordinate. This includes GRENOUILLE [8], STRIPED FISH [9], or biprism-based autocorrelators, all of which utilize a camera [10]. Next, spectral interferometry techniques such as SEA-SPIDER [11] or self-referenced spectral interferometry (SRSI) [12] retrieve the spectral phase directly, whereas dispersion-based methods, like d-scan [13] and its scan-free variants, determine pulse compression through controlled dispersion. Other concepts, including the amplitude-swing approach [13], further expand the available techniques to cite a few. However, these techniques often come at the expense of increased optical complexity and the need for detector arrays or cameras, making them most suitable for detailed measurements of complex or few-cycle pulses, preferentially at shorter wavelengths. Therefore, an open question remains of how much information about the pulse can be obtained solely from a single-pixel detector directly illuminated by the laser. Despite its apparent simplicity, a single-pixel detector can, in principle, enable shot-to-shot analysis, as the photocurrent response directly reflects instantaneous pulse-to-pulse fluctuations in power or duration when sampled at sufficient bandwidth. In the present work, however, we focus on time-averaged pulse characterization, as our goal is not fast real-time analysis but demonstrating the feasibility and accuracy of a compact, calibration-based diagnostic concept.

In this work, we use a common technological configuration: a packaged semiconductor LD integrated with a monitoring (MON) photodiode, which allows simultaneous measurement of linear and nonlinear responses, offering a compact platform for ultrafast pulse diagnostics. The use of a nonlinear detector based on TPA detection to retrieve pulse width in a moving-part-free setup has been previously demonstrated [14]. However, pulse duration estimation was possible by assuming an inverse dependence on peak power for a fixed pulse shape, repetition rate, and energy. In our case, using a miniature semiconductor laser with an integrated power-monitoring photodiode eliminates the need to maintain a constant pulse energy, enabling measurements within a single device and thus simplifying the alignment process.

The linear and nonlinear detectors were calibrated here by analyzing the dependence of both photocurrents on average and peak optical power, which enabled us to determine the pulse widths (the experiments were conducted with lasers delivering pulses in the 55–270 fs range). This calibration procedure is effective not only in free-space (FS) experimental setups but also when the semiconductor laser, serving as a detector, is coupled to an optical fiber. Once performed, pulse widths can be determined without relying on any moving parts—a practical advantage that greatly simplifies the measurement. Together, these results establish a simple and robust scheme for femto- or picosecond pulse characterization without moving parts.

## 2. Simultaneous linear and nonlinear detection of laser pulses

The Fabry–Pérot (FP) InGaAsP semiconductor laser (NEC, NX7302AA-CC, Fig. 1(a)–(c)) used in our experiment incorporates two semiconductor devices in one package. The first one – LD – has a nominal emission wavelength of 1356 nm and provides a quadratic response to intensity at longer wavelengths. The second one – MON – is integrated with the back facet of the FP laser [Fig. 1(c)], where it linearly detects a small fraction of light that passes through the partially transmitting back mirror [15] along with scattered light uncoupled to the waveguide. Considering a typical responsivity of InGaAs photodiodes at 1.5 μm approaching 1 A/W, in our setup we anticipate ~2% of free-space coupling efficiency and 0.5% in the fiber-coupled (FC) configuration (see Section 5 for details).

Unlike for the MON, the nonlinear signal generated in the LD originates from a third-order nonlinear process ($\chi^{(3)}$) TPA [16], which does not require phase matching or precise polarization control, greatly simplifying experimental implementation. Semiconductor LDs are particularly attractive as TPA detectors because they are commercially available, easy to integrate, and exhibit enhanced photocurrent due to their waveguide [15, 16]. The ability to obtain both linear and nonlinear signals within a single device offers several practical advantages.

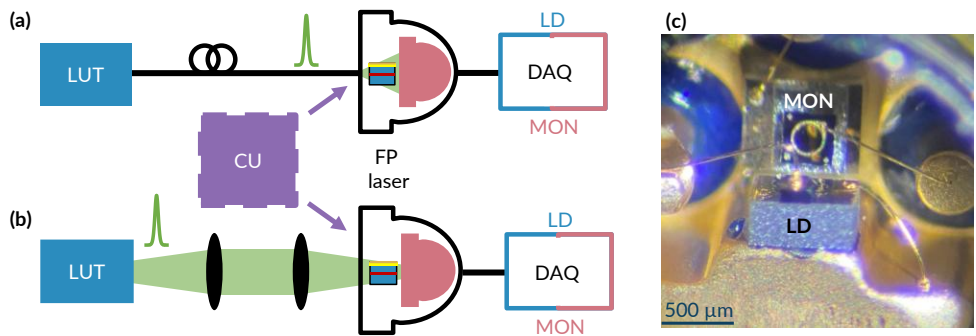

**Fig. 1.** Simultaneous nonlinear and linear detection in a InGaAsP FP LD. Conceptual detection scheme in (a) fiber-coupled (b) free-space configuration. An incident femtosecond pulse is detected simultaneously in a single housing without moving parts. LUT – laser under test, CU – calibration unit, DAQ – digital acquisition. (c) Microscope image of a LD chip and MON. For the free-space configuration, more light reaches the MON detector.

Collecting both signals simultaneously simplifies the experimental setup, reduces alignment requirements, and thus highlights the versatility of commercial FP LDs as compact, multifunctional diagnostic tools. In pulse width retrieval, the LD provides information on both pulse energy and duration; however, previous implementations have assumed a constant pulse energy. Including a MON to measure the average power removes this constraint.

## 3. Pulse duration estimation

Considering the absence of scanning and the ultrafast timescales involved, direct optoelectronic detection is not feasible. Therefore, to retrieve the pulse width from the average and peak values obtained with the FP laser, calibration against a reference unit is required [Fig. 1(a), (b)]. In linear detectors, the photocurrent is directly measured, as it scales linearly with incident optical intensity. This is here provided by the MON photodiode. However, since the FP laser acting as detector operates via TPA, its response is proportional to the square of optical intensity. Therefore, a squared LD photocurrent scales with peak optical power proportional to pulse energy and inversely proportional to pulse duration.

Photocurrents generated simultaneously by the LD and the MON in both FS and FC configurations were conditioned by a transimpedance amplifier (FEMTO DHPCA-100) and measured using a multimeter (Sanwa PC5000). An Er-doped, mode-locked fiber laser operating at a 250 MHz repetition rate and centered at 1550 nm served as the light source. The laser under test (LUT) provided tunable pulse durations from sub-55 fs to 270 fs, achieved through nonlinear amplification and spectral broadening. The pulse width was adjusted by varying the LUT pump current, while the average power was further controlled using neutral density (ND) filters at fixed pump conditions. To couple infrared light to the LD and MON detectors, a collimated beam with a $1/e^2$ diameter of 3.6 mm was focused by a ½-inch 15 mm focal length off-axis parabolic mirror. Figure 2(a) shows the square root of the LD detector response when both pulse width and optical power were varied, corresponding to changes in peak intensity. A clear linear dependence is observed across the full scan range. In contrast, the MON detector exhibits a linear response to the average power [Fig. 2(b)].

Pulse durations necessary for the calibration were characterized by recording fringe-resolved intensity autocorrelations (FRAC, Fig. 3(a)–(c)), intensity autocorrelations (IAC, Fig. 3(d)–(f)). Although this experiment leveraged a dispersion-compensated Michelson interferometer (described and calibrated as in Ref. [18]) with the conventional detector replaced by the InGaAsP FP LD with MON, pulse diagnostics following the one-time calibration is moving-parts free.

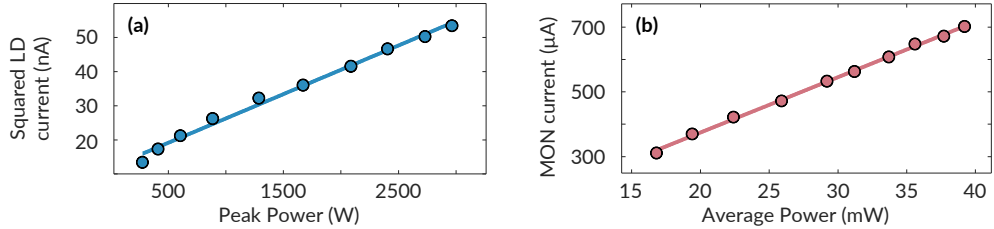

**Fig. 2** (a) Peak power as a function of squared LD current. (b) Average power as a function of MON current. Both curves reveal linear dependence.

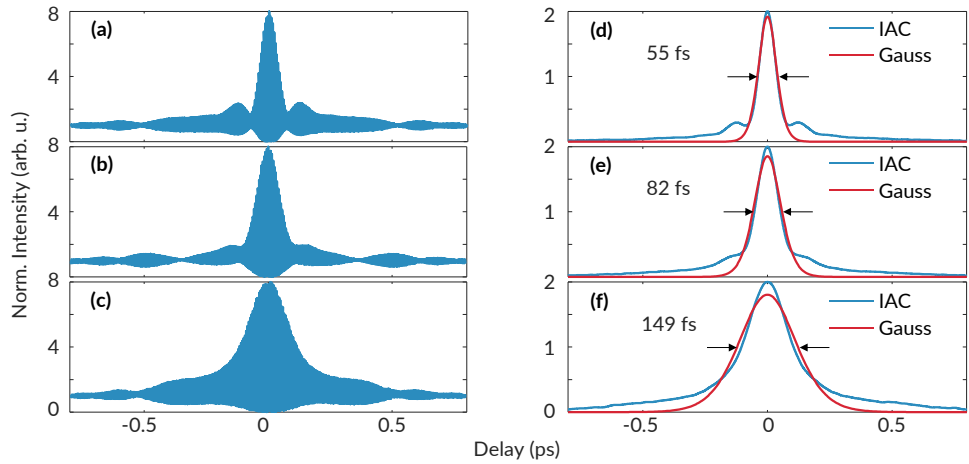

**Fig. 3.** FRAC, IAC traces obtained via LD based autocorrelator. (a)–(c) FRAC traces, (d)–(f) IAC traces with 55 – 149 fs pulse duration.

The recorded IAC traces were fitted with a Gaussian fit [Fig. 3(d)–(f)], and a 1.41 deconvolution factor was applied to retrieve the corresponding pulse duration. Next, it can be linked to the measured photocurrents by analyzing the ratio of the linear MON photocurrent - $I_{\mathrm{MON}}$, to the squared LD photocurrent – $I_{\mathrm{LD}}$. The nonlinear response of the latter, dominated by TPA, scales with the peak power of the pulse:

$$\sqrt{I_{\mathrm{LD}}} \sim P_{\mathrm{peak}} \approx 0.94 \frac{P_{\mathrm{av}}}{f_{\mathrm{rep}} \tau_{\mathrm{p}}}, \qquad (1)$$

where $P_{\mathrm{av}}$ is the average power, $f_{\mathrm{rep}}$ is the laser repetition rate, $\tau$ is the pulse duration and 0.94 is a constant factor for Gaussian-shaped pulses. To eliminate the constraint of a fixed average power, the linear current was divided by the nonlinear current to obtain a photocurrent ratio:

$$R = \frac{I_{\mathrm{MON}}}{\sqrt{I_{\mathrm{LD}}}} = \frac{1}{f_{\mathrm{rep}} \tau_{\mathrm{p}}}. \qquad (2)$$

The established calibration curve $R(\tau_p)$ [Fig. 4(b)] captures the dependence of $1/\tau_{\mathrm{p}}$ on the photocurrent ratio, enabling the reconstruction of unknown pulse widths from measured photocurrents. Once this relationship is determined, the pulse duration of any subsequent pulse can be directly inferred from the measured linear and nonlinear photocurrents, without requiring any moving parts or interferometric scanning.

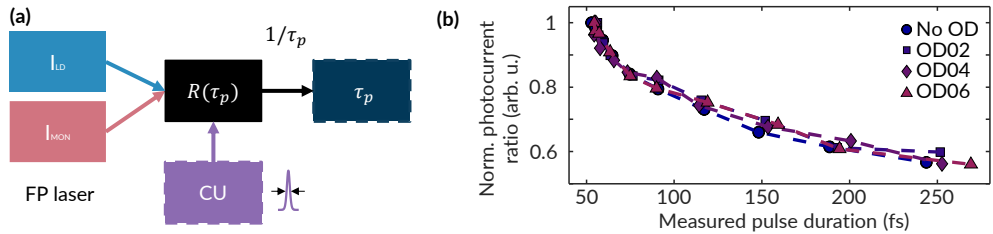

**Fig. 4.** Pulse duration retrieval calibration. (a) Conceptual calibration scheme, (b) calibration curves based on measured photocurrents ratio for various optical intensities. OD02 corresponds to an optical density of 0.2 (2 dB of attenuation), while OD04 and OD06 correspond to attenuations of 4 dB and 6 dB, respectively.

We want to emphasize an important practical advantage of this technique. The normalized photocurrent ratio *R does not* depend on average power (here modified by ND filters in steps of 2 dB to 6 dB) for a given pulse duration, as shown in Fig. 4(b). This establishes the reproducibility and replicability of the calibration method.

## 4. Pulse duration retrieval via simultaneous detection in free space

We retrieved the pulse duration using the calibration curve [Fig. 4(b)] based on the established relationships for both LD and MON responses [Fig. 2(a), (b)]. The reconstructed pulse durations [Fig. 5(a)] had a root mean square error of prediction (RMSEP) of 4.41 fs and a coefficient of determination $R^2 = 0.995$. The most significant discrepancies were observed for shorter pulses in the 50–120 fs range, with residuals of −3.3 fs at 53 fs (6.3% relative error) and 5.8 fs at 75 fs (–7.8% relative error) [Fig. 5(b), (d)].

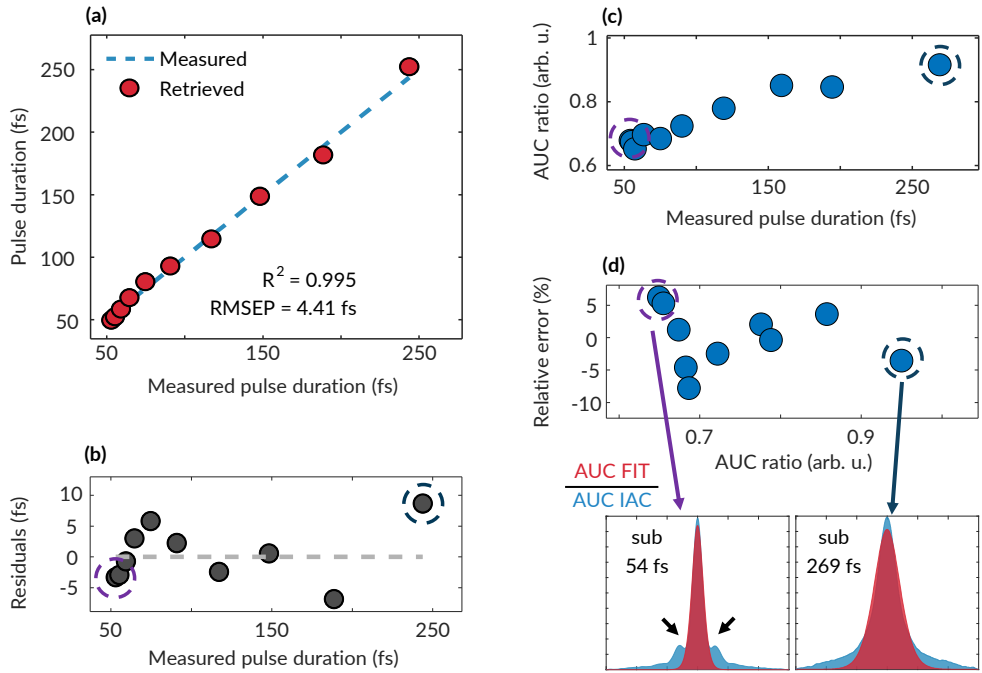

**Fig. 5.** Pulse duration retrieval via simultaneous nonlinear and linear photodetection. (a) Retrieved pulse duration vs one measured using an optical autocorrelator. (b) Residuals of retrieved pulse durations vs measured ones. (c) Area under curve ratio of a Gaussian fit to IAC data as a function of measured pulse duration. It is clear that the shortest pulses had pronounced wings in the IAC traces, which signify an oscillating pulse tail. Such pulses were not Fourier-limited. (d) Relative error as a function of the area under the curve ratio. The relative error decreased as AUC increased. Incorporating this coefficient into the analysis, therefore, enables more accurate prediction and correction of the retrieved pulse widths.

These deviations arise from the intrinsic nature of bandwidth-limited photocurrent-based measurements, which capture only the average value while being sensitive to the pulse shape. Additional inaccuracies stem from assuming a Gaussian fit to the IAC trace, which can further distort the extracted width. To mitigate these effects, we introduce an area under the curve (AUC) ratio, defined as the ratio of the AUC of the fit to that of the IAC:

$$r_{\text{AUC}} = \frac{\int_{-\tau_{\min}}^{\tau_{\max}} A_{\text{FIT}}^{(2)}(\tau)\,d\tau}{\int_{-\tau_{\min}}^{\tau_{\max}} A_{\text{M}}^{(2)}(\tau)\,d\tau}. \quad (3)$$

For broader pulses with reduced chirp, the AUC ratio approaches 0.9, resulting in a relative error as low as –3.5% [Fig. 5(d)].

Once calibrated, the presented pulse retrieval routine provides a simple alternative to conventional autocorrelation methods, as it eliminates the need for moving parts and complex alignment. The main discrepancies obviously arise for ultrashort, chirped pulses, where the photocurrent measurement captures only time-averaged intensity. By incorporating the AUC ratio into the analysis, these limitations can be anticipated.

## 5. Pulse duration retrieval in fiber-coupled detector

Previous experiments were performed with the detector in a FS configuration [Fig. 1(b)]. However, fiber-coupled or fiber-compatible operation [Fig. 1(a)] is of practical relevance because it inherently resists misalignment and eliminates the need for external focusing. It also enables easy integration with fiber-based setups, demonstrating that the method performs consistently regardless of the light coupling scheme. This demand in our experiment was addressed by an analogous FP laser diode with (Luminent, MRLDFC010-NEB1) with an integrated fiber connector.

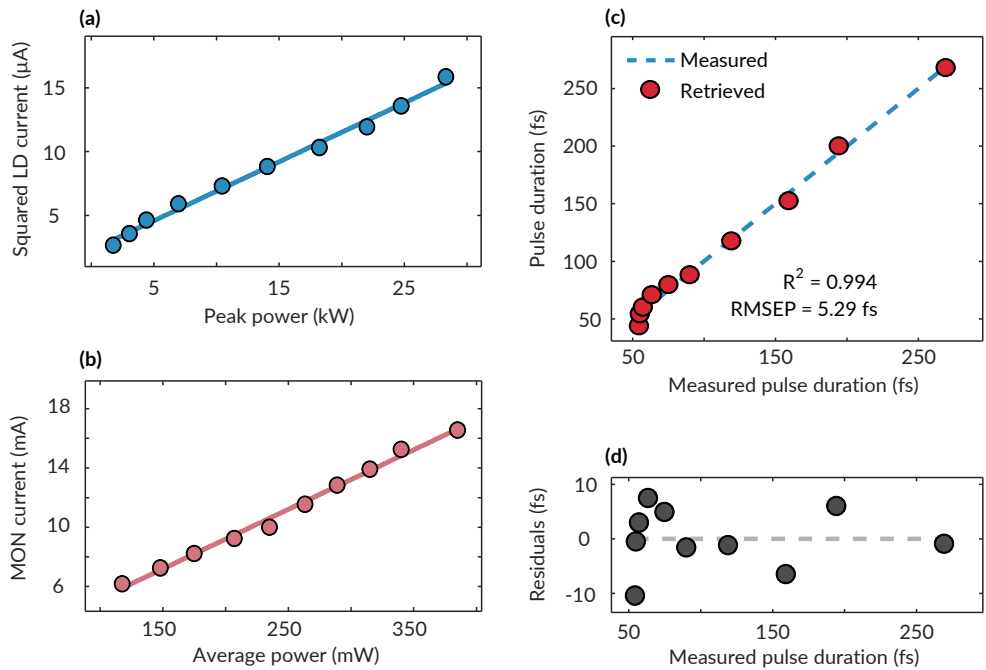

**Fig. 6.** Pulse duration retrieval in fiber coupled detector (a) Peak power as function of squared LD current (b) Average power as function of MON current. (c) Retrieved pulse durations of FC configuration. (d) Residuals.

The measured photocurrent dependencies shown in Fig. 6(a), (b) remained linear, confirming the validity and repeatability of the calibration In the FC case, pulse widths were retrieved [Fig. 6(c)] with a coefficient of determination $R^2 = 0.994$ and a root-mean-square error of prediction (RMSEP) of 5.29 fs, slightly higher than in the FS setup. The corresponding residuals [Fig. 6(d)] ranged from –10.4 fs to 7.5 fs, consistent with the expected variation due to pulse shape and subsequent oscillations of the main pulse. Despite the higher average (400 mW) and peak (30 kW) powers used, pulse width retrieval remains feasible for average powers as low as 1.9 µW, where the signal-to-noise ratio was 2, owing to the waveguide-enhanced photocurrent response of the LD. Once calibrated, both configurations can be used to retrieve femtosecond pulses with the photocurrent-based detection scheme. FC enables integration with fiber-based systems, while FS measurements offer minimal dispersion for shorter pulses, together validating the method's simplicity, robustness, broad applicability, and reliability regardless of how the incident light is delivered. Reproducibility of the method is also an important factor to highlight, as for both experiments, different LDs were used, however, with an identical semiconductor material system (InGaAsP).

## 6. Conclusion

We have demonstrated that commercially available packaged Fabry-Pérot laser diodes with monitoring photodiodes can function as photodetectors that simultaneously measure linear and nonlinear photocurrents. Probing them from an InGaAsP laser diode and its integrated InGaAs monitor photodiode, we determined the ratio of the average optical power to the pulse peak intensity, which enabled us to retrieve pulse widths from ~55 to 270 fs without moving parts with single fs accuracy. Despite the requirement for initial calibration, this approach allows for unambiguous pulse width measurement even when the average power varies. This approach should also enable pulse width retrieval across a wider duration range via extrapolation (i.e. to the picosecond range). When the optical pulse is not clean and exhibits significant oscillations following the main pulse, more significant estimation errors may occur. We found that an analysis of the area under curve ratio between a Gaussian fit and the measured autocorrelation profile provides a means to estimate the prediction error and its eventual reduction.

A comparison between free space and fiber coupled configurations further revealed that both can be employed reliably once calibrated: the free space coupled enables higher flexibility in alignment, while the fiber connectorized one facilitates integration into all-fiber and compact systems. This simple, low-power-compatible, and robust approach provides a practical, straightforward alternative to more complex diagnostic methods with potential applications extending to dual-comb ranging and nonlinear imaging systems.

**Funding.** Narodowe Centrum Nauki (2022/45/B/ST7/03316);

**Acknowledgments.** We thank Jakub Mnich for the discussion and construction of the optical delay line.

This work is supported by the use of the National Laboratory for Photonics and Quantum Technologies (NPLQT) infrastructure, which is financed by the European Funds under the Smart Growth Operational Programme. S. A.

**Disclosures.** The authors declare no conflicts of interest.

**Data availability.** Data underlying the results presented in this paper are available in Ref. [19].